\documentclass[journal]{IEEEtran}
\usepackage{cite}
\usepackage{graphicx}
\usepackage{amsmath,amssymb,amsfonts}
\usepackage{textcomp}
\usepackage{xcolor}
\usepackage{booktabs}
\usepackage{array}
\usepackage{siunitx}
\usepackage{url}
\usepackage{hyperref}
\hypersetup{colorlinks=true,linkcolor=black,citecolor=black,urlcolor=blue}
\usepackage{pgfplots}
\pgfplotsset{compat=1.18}
\usepgfplotslibrary{groupplots}
\usetikzlibrary{arrows.meta,matrix}

\usepackage{titlesec}
\titlespacing*{\section}{0pt}{6pt}{3pt}

\usepackage{titlesec}
\titlespacing*{\subsection}{0pt}{6pt}{3pt}

\begin{document}

\title{SYNTLOG:\\ FSM Benchmarks Evaluation for FPGA}

\author{Samary~Baranov and~Danila~Gorodecky
\thanks{S.\ Baranov is with Holon Institute of Technology, Holon, Israel
(e-mail: samarybaranov@gmail.com).}%
\thanks{D.\ Gorodecky is with INESC-ID, Instituto Superior T\'ecnico,
Universidade de Lisboa, Lisbon, Portugal (e-mail: danila.gorodecky@gmail.com).}}

\markboth{IEEE Embedded Systems Letters,~Vol.~XX, No.~X, Month~2026}%
{Shell \MakeLowercase{\textit{et al.}}: A Sample Article Using IEEEtran.cls for IEEE Journals}


\maketitle

\begin{abstract}
We introduce a curated benchmark collection of \num{101} FSM descriptions organized into five size
classes (\emph{small}, \emph{medium}, \emph{large}, \emph{huge}, \emph{super-huge}), spanning tens to thousands of states and up to hundreds of inputs and outputs. Using these benchmarks we compare an architecture-independent synthesis tool SYNTLOG against Xilinx Vivado at the synthesis (technology-mapping) level, under six optimization strategies. We compare three post-synthesis parameters: number of LUTs, number of logic levels, and runtime. SYNTLOG in area-driven mode uses substantially fewer LUTs than Vivado's area baseline at every size ($-45\%$ to $-66\%$), and its delay-driven mode is simultaneously shallower in logic depth than Vivado's default synthesis mode. SYNTLOG synthesis includes embedded functional validation. Runtime of SYNTLOG is faster by one to two orders of magnitude than Vivado synthesis. The proposed tool synthesizes and validates every \emph{huge} and \emph{super-huge} circuit in minutes, whereas Vivado fails to terminate on these designs after hours.
\end{abstract}


\section{Introduction}
Finite state machines (FSMs) are the backbone of control logic in digital systems, yet experimental work on their minimization is thin. Theory on FSM assignment, structural decomposition, and
ASM-based (Algorithmic State Machine) design is well established \cite{baranov2018,devadas}, and dedicated state-encoding algorithms such as NOVA~\cite{nova} and LUT-oriented FSM synthesis methods~\cite{czerwinski} have been studied for decades. Yet reproducible empirical evaluation remains constrained by available data. The widely cited
suites ITC'99~\cite{itc99}, the MCNC/LGSynth FSM set, and circuits bundled with academic tools such as ABC~\cite{abc} and SIS~\cite{sis}, contain few examples, most of them small. As a result, claims about synthesis behavior on large, control-dominated designs are rarely supported by measurement.

This paper makes two contributions. First, we describe a benchmark collection broader and larger than existing FSM suites, organized into five size classes so that scaling behavior is directly observable. Second, we use it to compare an architecture-independent synthesis tool SYNTLOG with the Xilinx  Vivado flow across optimization strategies. Because sizeable FSM benchmarks are scarce, results on such a collection are of independent value to other researchers.

Section~\ref{sec:bench} presents the collection; Section~\ref{sec:tool}
summarizes the synthesis approach; Section~\ref{sec:method} details the
experimental setup; Section~\ref{sec:results} reports and discusses results; and
Section~\ref{sec:concl} concludes.

\section{Benchmark Collection}\label{sec:bench}
The benchmarks are drawn from a recent public collection of FSM descriptions \cite{benchsite}. Existing suites are limited in two ways at once: few designs and small designs. The present collection addresses both, with \num{101} FSM descriptions in HDL grouped by size into five classes. Table~\ref{tab:params} gives, per class, the number of designs and the average number of inputs, outputs, states, and number of conditions in the transition  table.

\setlength{\tabcolsep}{0.2em}
\renewcommand{\arraystretch}{1.15}
\begin{table}[!t]
\caption{FSM benchmark groups and average parameters.}
\label{tab:params}
\centering
\setlength{\tabcolsep}{4pt}
\begin{tabular}{l c c c c c}
\toprule
Group & \# of benchmarks & \multicolumn{4}{c}{Average}\\
\cmidrule(l){3-6}
 & per group & Inputs & Outputs & States & Lines\\
\midrule
Small      & 41 & 14.5 & 24.2 & 20.2  & 84\\
Medium     & 44 & 21.8 & 50.4 & 58.6  & 247\\
Large      & 6  & 54.7 & 58.7 & 138.3 & 1\,121\\
Huge       & 7  & 64.1 & 66.1 & 627.7 & 24\,114\\
Super-Huge & 3  & 76.7 & 97.0 & 2\,309.7 & 206\,062\\
\bottomrule
\end{tabular}
\end{table}

The dynamic range of \emph{super-huge} averages more than \num{2300} states and \num{206000} transitions. FST compiled from regular-expression rule sets for hardware-based deep packet inspection~(DPI) in network intrusion detection systems~(NIDS) routinely reach tens of thousands of states~\cite{dpi_approx,odrem}. FSMs serve as control units in processors, protocol controllers, digital signal processing~(DSP) blocks, and neural-network activation circuits~\cite{barkalov2024}. These designs lie well beyond any classical FSM benchmark and provide a stress test that separates scalable synthesis from heuristic flows that flatten or restructure logic.


\section{Architecture-Independent Synthesis}\label{sec:tool}
SYNTLOG adopts a future-proof view of design. It minimizes FSMs for both
ASIC and FPGA targets and is not tied to a fixed device architecture. For FPGAs
it generates RTL for look-up tables (LUTs) of an arbitrary input width $M$, so the same description
can be retargeted to new device generations without reengineering, while Xilinx Vivado flows, by contrast, are tightly coupled to today's 4-, 5-, or 6-input LUTs. Since our
reference is the current Xilinx FPGA family, whose logic cell is a 6-input LUT,
all results in this letter are reported for $M=6$. For ASIC targets the tool
realizes the control-unit FSM as a regular matrix structure.

SYNTLOG is also flexible in how a machine is specified and encoded. An FSM may
be supplied as a state-transition table (a list of transitions), as an
ASM, or as an HDL description (VHDL or Verilog), and
its states may be encoded either as one-hot (a single asserted bit per
state) or in compact binary form (using $\lceil\log_2 S\rceil$
flip-flops for $S$ states). Both area- and delay-driven optimization are
supported, and beyond FSMs the same engine also minimizes purely combinational
circuits. Throughout this study we use one-hot encoding at $M=6$ to match the
target FPGAs. Rather than flattening the logic, the tool synthesizes structured
RTL directly from the control-flow graph, using look-ahead transition logic so
that even deeply nested transitions remain compact, following the ASM-based
design methodology \cite{baranov2018}. This structure is what lets it scale where
heuristic, flattening-based tools fail.

A distinctive feature is embedded validation. The tool does not only emit RTL, it
performs a behavioral simulation that exercises every state and transition
of the machine and checks the synthesized circuit's response against the source
FSM, confirming exact functional equivalence rather than sampling a subset of the
input space. Every runtime reported below includes this
simulation step. The approach scales to very large machines: the largest example
in our collection has 2314 states and 351351 transitions, for which
fully validated RTL was produced in under a half hour including simulation.

\section{Experimental Setup}\label{sec:method}
We conducted experiments used Xilinx Vivado~2022.2 on Kintex-7 (\texttt{xc7k70tfbg484-3}) and Kintex~UltraScale+ (\texttt{xcku5p-ffvb676-2-e}), both built around 6-input LUTs. Each benchmark was synthesized as a standalone top module. To keep the comparison confined to the LUT/flip-flop, fabric block-RAM inference was disabled, so that FSM logic maps to LUTs and registers only.

We compare results at the synthesis (technology-mapping) level, before place-and-route. SYNTLOG emits vendor-neutral RTL rather than a placed-and-routed implementation, and synthesis already fixes the LUT network, register count, and logic depth that we report.

The two device families behave almost identically in LUTs, but not in logic levels. Post-synthesis LUT counts differed by less than $1\%$ between Kintex-7 and Kintex~US+ across all groups and strategies. The LUT result depends on the LUT inputs $M$, but not on the specific device, recisely the parameter that the proposed tool targets. Logic depth, however, can differ by as much as $25\%$ for the same optimization strategy, since timing-driven mapping relies on device-specific delay models. Because we do not analyze individual circuits but average each of the four reported metrics (LUT count, logic depth, runtime) over a whole benchmark group, such per-example differences have little effect on the group means. Therefore we report Kintex-7 figures only, and only for the three strategies whose results differ most: \emph{Area\_Default} (targeting minimal LUT count), \emph{Speed\_200\,MHz\_Performance} (a balanced trade-off between LUT count and logic depth), and \emph{Speed\_1\,GHz\_Performance} (targeting minimal logic depth).

Table~\ref{tab:strats} shows 6 Vivado optimization strategies during synthesis\footnote{
The results for all optimization strategies and for both FPGA families in \url{https://github.com/ZeboZebo702/SYNTLOG_Benchmarks_FPGA}}. \emph{Area\_Default} applies no timing constraint and the default flow; \emph{Balanced\_100\,MHz} adds a relaxed \SI{10}{\nano\second} period; the \emph{Speed\_200\,MHz} pair tightens it to \SI{5}{\nano\second}, once with the default flow and once with performance-oriented directives (\texttt{Flow\_PerfOptimized\_high}, \texttt{Performance\_Explore}) that trade area for shorter critical paths; the \emph{Speed\_1\,GHz} pair sets an aggressive \SI{1}{\nano\second} period. 
For SYNTLOG we report results at the matching LUT width $M=6$ under both area-
and delay-driven optimization. 
\setlength{\tabcolsep}{0.2em}
\renewcommand{\arraystretch}{1.15}
\begin{table}[!t]
\caption{Six applied Vivado optimization strategies.\\ Boldface: the three strategies reported
in Section~\ref{sec:results}.}
\label{tab:strats}
\centering
\setlength{\tabcolsep}{3.5pt}
\begin{tabular}{l c l}
\toprule
Strategy & Period & Synthesis directive\\
\midrule
\textbf{Area\_Default}           & --                    & Vivado defaults\\
Balanced\_100\,MHz               & \SI{10}{\nano\second} & Vivado defaults\\
Speed\_200\,MHz Default          & \SI{5}{\nano\second}  & Vivado defaults\\
\textbf{Speed\_200\,MHz Perf.}   & \SI{5}{\nano\second}  & PerfOptimized\_high\\
Speed\_1\,GHz Default            & \SI{1}{\nano\second}  & Vivado defaults\\
\textbf{Speed\_1\,GHz Perf.}     & \SI{1}{\nano\second}  & PerfOptimized\_high\\
\bottomrule
\end{tabular}
\end{table}


\section{Results and Discussion}\label{sec:results}
Table~\ref{tab:main} reports, for every benchmark group, the average LUT count,
logic depth (LUT levels), flip-flop (FF) count, and runtime for three Vivado strategies on
Kintex-7 \emph{Area\_Default} (AD), \emph{Speed\_200\,MHz\_Performance} (200P) and
\emph{Speed\_1\,GHz\_Performance} (1GP), and for SYNTLOG at $M=6$ under
area-driven (SLG\_A) and delay-driven (SLG\_D) optimization. Vivado completed the
\emph{small}, \emph{medium} and \emph{large} groups, but did not terminate on any
\emph{huge} or \emph{super-huge} design, those cells are blank.

\setlength{\tabcolsep}{0.2em}
\renewcommand{\arraystretch}{1.15}
\begin{table*}[!t]
\caption{Average post-synthesis results on Kintex-7. Vivado
\emph{Area\_Default} (\emph{AD}), \emph{Speed\_200\,MHz\_Performance} (\emph{200P}), \emph{Speed\_1\,GHz\_Performance} (\emph{1GP}). SYNTLOG (for $M=6$) \emph{SLG\_A} is area-driven and \emph{SLG\_D} is delay-driven, whose flip-flops (\emph{FF}) count is one-hot ($=$ number of states) and identical for both objectives (\emph{SLG}). ``--'' marks designs Vivado did not synthesize.
SYNTLOG runtime includes functional validation; Vivado times are synthesis only.}
\label{tab:main}
\centering
\scriptsize
\setlength{\tabcolsep}{2.4pt}
\begin{tabular}{l | rrrrr | rrrrr | rrrr | rrrrr}
\toprule
& \multicolumn{5}{c}{Average LUTs} & \multicolumn{5}{c}{Average logic depth}
& \multicolumn{4}{c}{Average FF} & \multicolumn{5}{c}{Average runtim, \textit{s}}\\
\cmidrule(lr){2-6}\cmidrule(lr){7-11}\cmidrule(lr){12-15}\cmidrule(lr){16-20}
Group
& AD & 200P & 1GP & SLG\_A & SLG\_D
& AD & 200P & 1GP & SLG\_A & SLG\_D
& AD & 200P & 1GP & SLG
& AD & 200P & 1GP & SLG\_A & SLG\_D\\
\midrule
Small      & 98   & 56   & 63   & 54   & 70   & 5.2 & 2.3 & 2.4 & 5.4  & 3.9  & 10 & 22 & 23 & 20   & 38  & 45 & 45 & 0.3 & 0.3\\
Medium     & 288  & 133  & 146  & 123  & 165  & 5.9 & 2.9 & 3.1 & 7.3  & 5.1  & 10 & 64 & 68 & 59   & 43  & 49 & 49 & 0.3 & 0.4\\
Large      & 2023 & 1224 & 1304 & 696  & 941  & 8.0 & 3.2 & 4.2 & 10.0 & 6.7  & 70 & 141& 148& 138  & 161 & 86 & 88 & 0.6 & 1.0\\
Huge       & --   & --   & --   & 7585 & 9949 & --  & --  & --  & 18.0 & 10.6 & -- & -- & -- & 628  & --  & -- & -- & 40  & 48\\
Super-Huge & --   & --   & --   & 32\,380 & 37\,718 & -- & -- & -- & 27.0 & 15.3 & -- & -- & -- & 2310 & -- & -- & -- & 1014 & 1177\\
\bottomrule
\end{tabular}
\end{table*}

\subsection{State encoding}
The FF column is the key to reading the rest of the table. The proposed tool uses
\emph{one-hot} encoding throughout, i.e. FF count equals the number of states exactly,
e.g.\ the \emph{super-huge} "hippopotamus" design has 2314 states and
2314 flip-flops, and is identical for area- and delay-driven
optimization. Encoding is thus a fixed structural property of the tool. Vivado, by contrast, changes encoding with the strategy: an FF count cannot change by combinational optimization alone, yet Vivado's count grows $2\times$--$7\times$ from \emph{Area\_Default} to the performance strategies
(e.g. $10\rightarrow 68$ on \emph{medium} and $70\rightarrow 148$ on \emph{large}). \emph{Area\_Default}
uses a compact (binary/sequential) encoding, whereas 200P and 1GP switch to one-hot. The
like-for-like comparison to the proposed one-hot tool is therefore Vivado's
performance strategies, which we keep in view alongside the area baseline below.

\subsection{Mapped area (LUTs)}
The proposed tool is markedly LUT-efficient. Against Vivado's area baseline
(\emph{Area\_Default}, a compact encoding) SYNTLOG's area-driven mode uses far fewer
LUTs at every size: $54$ vs.\ $98$ on \emph{small} ($-45\%$), $123$ vs.\ $288$ on
\emph{medium} ($-57\%$), and $696$ vs.\ \num{2023} on \emph{large} ($-66\%$). That a
one-hot design undercuts Vivado's compact encoding on area is notable:
the structured, ASM-derived next-state logic more than compensates for the larger
state register. At matched one-hot encoding (against 200P) the picture is just as
favorable on the larger designs: the tool's area mode ties on \emph{small} ($54$ vs.\
$56$), wins on \emph{medium} ($123$ vs.\ $133$) and wins decisively on \emph{large}
($696$ vs.\ \num{1224}, $-43\%$). The LUT advantage therefore holds across both the
area- and the performance-oriented Vivado configurations, and widens with design size.

\subsection{Logic depth}
SYNTLOG's delay-driven mode is shallower than Vivado \emph{Area\_Default} at every completed size ($3.9$/$5.1$/$6.7$ vs.\ $5.2$/$5.9$/$8.0$ levels), and trails only Vivado's aggressively timing-driven one-hot strategies ($2.3$--$3.2$ levels), at the same time as using fewer LUTs than either on \emph{large}. Within the tool, switching from area- to delay-driven optimization lowers depth substantially at the cost of LUTs, while keeping the same FF count.


\subsection{Runtime}
The runtime difference is the most pronounced. On the groups Vivado completes the proposed
tool is faster by roughly two orders of magnitude: \SI{0.3}{\second} vs.\
\SIrange{38}{45}{\second} on \emph{small}, \SI{0.3}{\second} vs.\ \SIrange{43}{49}{\second}
on \emph{medium}, and \SIrange{0.6}{1.0}{\second} vs.\ \SIrange{86}{161}{\second} on
\emph{large}. The proposed-tool figure already includes functional validation, while the
Vivado figure is synthesis only (no place-and-route), so the end-to-end gap is larger
still.

\subsection{Scalability}\label{sec:scal}
The decisive result is qualitative. None of the three \emph{super-huge} benchmarks, and several \emph{huge} benchmarks, could be synthesized by Vivado: after tens of hours the process was aborted without a result. The proposed tool synthesized and validated every one of them (\SI{40}{\second} (area) on average for \emph{huge} and $\approx$\SI{17}{\minute} for \emph{super-huge}) with the expected one-hot FF counts. 

\subsection{Device and architecture independence}
Because Kintex-7 and Kintex~UltraScale+ both implement 6-input LUTs, their post-synthesis
LUT counts differ by under $1\%$; the device is immaterial at this level, and the relevant
parameter is the LUT width $M$, which SYNTLOG targets directly. The results above
use $M=6$ to match the FPGAs, but the same one-hot description can be retargeted to any width
without redesign, which is the tool's principal architectural advantage and the subject of
ongoing experiments.

\section{Conclusion}\label{sec:concl}
We presented a five-class FSM benchmark collection, larger and broader than
existing public suites, and used it to compare the architecture-independent,
ASM/FSM-driven tool SYNTLOG with Xilinx Vivado~2022.2 at the synthesis level. For FPGA, SYNTLOG encodes states as one-hot, so its flip-flop count equals the number of states and is independent of the optimization objective. Vivado
switches encoding by strategy (compact for area, one-hot for performance), which is why its register count varies $2\times$--$7\times$. SYNTLOG is consistently more LUT-efficient: its area-driven mode uses $45\%$--$66\%$ fewer LUTs than Vivado's area baseline at every size, and at matched one-hot encoding it ties on
small and wins on medium and large ($-43\%$). Its delay-driven mode is shallower than Vivado's default synthesis at every completed size and trails only Vivado's aggressively timing-driven strategies. SYNTLOG synthesizes one to two orders of magnitude faster than Vivado while validating each result by full behavioral simulation, and it completes and validates every huge and super-huge design on which Vivado does
not terminate. Because post-synthesis area depends on the LUT width $M$ rather than the device, the ability to target an arbitrary $M$ is the relevant degree of freedom for future architectures.

\end{document}